\documentclass[%
 reprint,
 amsmath,amssymb,
 aps,
]{revtex4-1}

\usepackage{graphicx}
\usepackage{dcolumn}
\usepackage{bm}
\usepackage{dsfont}
\usepackage{xcolor}
\usepackage{ulem}

\def\m{M_{\rm QCD}}

\begin{document}

\preprint{APS/123-QED}
\title{Universal scale factors relating mesonic fields and quark operators}
\author{Amir H. Fariborz}
\email{fariboa@sunypoly.edu}
\affiliation{Department of Mathematics/Physics,  SUNY Polytechnic Institute, Utica, NY 13502, U.S.A.
}
\author{J. Ho}
\email{J.Ho@usask.ca}
\affiliation{%
	Department of Physics and
	Engineering Physics, University of Saskatchewan, Saskatoon, SK,
	S7N 5E2, Canada
}%
\author{T.G. Steele}
\email{Tom.Steele@usask.ca}
\affiliation{%
	Department of Physics and
	Engineering Physics, University of Saskatchewan, Saskatoon, SK,
	S7N 5E2, Canada
}%

\begin{abstract}
Scale factor matrices relating mesonic fields in chiral Lagrangians and quark-level operators of QCD sum-rules are shown to be constrained by chiral symmetry, resulting in universal scale factors for each chiral nonet.  
Built upon  this interplay between chiral Lagrangians and QCD sum-rules,   the scale factors relating the $a_0$  isotriplet scalar mesons to their underlying quark composite fields were recently determined.   It is shown that 
the same technique when applied to $K_0^*$ isodoublet scalars reproduces the same scale factors, confirming the universality property and
further validating   this connection between chiral Lagrangians and QCD sum-rules which can have nontrivial  impacts on our understanding of the low-energy QCD, in general, and the physics of scalar mesons in particular.   
\end{abstract}
\maketitle
\section{Introduction}
In the absence of an exact solution to the strong coupling limit of QCD in terms of fundamental quarks, replacing the fundamental degrees of freedom with light hadrons at low-energies has been shown to be a useful approximation in linear and nonlinear sigma model based approaches. Such 
approaches have also been proven to be very challenging; nevertheless, the great efforts by many investigators have led to significant progress over the past several decades and resulted in development of important frameworks such as chiral perturbation theory and various effective models \cite{PDG,PKZ_review}. Such frameworks parallel important properties of fundamental QCD by respecting several guiding principles such as chiral symmetry (and its breakdown), U(1)$_{\rm A}$ axial anomaly and various assumptions  about the QCD vacuum.     

However,  the quest for understanding the strong interaction phenomena at low-energies based on fundamental QCD has never stopped and important attempts have been made, most important of which is the approach of the lattice QCD program, which despite all the technical challenges has made an enormous progress \cite{PDG,Dudek:2013yja,Mathur:2006bs,Kunihiro:2003yj,Prelovsek:2004jp,McNeile:2006nv}. Still, a framework that can directly connect the low-energy strong interaction data to the fundamental quarks and gluons has not yet emerged. 
    Particularly in the scalar meson sector of low-energy QCD,  establishing such a connection is even less trivial.   On  the experimental side, some of these states are broad and overlap with nearby states, leaving some of their experimental properties vague.  On the theoretical side, explaining their mass spectrum and decay properties requires a description  beyond the conventional quark-antiquark pattern.  For the case of isosinglet scalars, the complexities are significantly greater because these states have the same quantum numbers of the QCD vacuum, which can develop a vacuum expectation value  and spontaneously break the chiral symmetry. This means that understanding the substructure of isosinglet scalars, which can be composed of not only various two- and four-quark fields but also of glue, is naturally nontrivial and perhaps beyond the current reach of lattice simulations.  For a full understanding of scalar sector, it is vital to seek a bridge that can connect the low-energy data all the way to fundamental QCD.  Such a solid bridge currently does not exist, and awaits the exact solution to nonperturbative QCD.   

In Refs.~\cite{Fariborz:2015vsa,Fariborz:2019zht}
 we demonstrated how a linkage between two existing frameworks, QCD sum-rules \cite{SVZ,Reinders:1984sr} (that significantly connect fundamental QCD to hadronic physics through duality relations) and chiral Lagrangians (which are appropriately designed in terms of the hadron fields and can be conveniently used to describe low-energy data) can provide an approximation of such a bridge.  
This linkage occurs through scale-factor matrices relating mesonic fields
of chiral Lagrangians to quark-level composite-operator current structures of QCD
sum-rules. In principle,  this connection is quite general and can be established with any formulation of QCD sum-rules and any type of chiral Lagrangian, but while the idea is general,  one naturally has to make a choice for each of these two frameworks based on their effectiveness in probing a particular channel or process.   Since our focus is on scalar channel  we use Gaussian sum-rules, a sum-rule methodology designed to handle hadronic mixing \cite{gauss,harnett_quark};
for the chiral Lagrangian side, we use the generalized linear sigma model of 
\cite{Fariborz:2015vsa,GLSM} which has been applied to a wide range of low-energy scattering and decay channels in which scalar mesons play a dominant role. Specifically, the scale factors were first determined for
the isovector scalar sector  \cite{Fariborz:2015vsa,Fariborz:2019zht} by connecting the QCD sum-rules to the chiral Lagrangian described by the generalized linear sigma model \cite{GLSM}. 
However, chiral symmetry requires that the scale factors must be universal for all members of the chiral nonets.  In the present work (using the same framework of Ref.~\cite{Fariborz:2015vsa,Fariborz:2019zht}), we demonstrate that the same scale factors 
are remarkably recovered in the isodoublet scalar sector, providing a crucial test of the universality property.

Establishing universality of these scale factors is essential for exploiting the bridge between chiral Lagrangians and QCD sum-rules to address the long-standing puzzles in the isoscalar sector. 
The exact  relationship between the composite fields of quarks representing a mesonic state (which requires a mass dimension of three or higher), and that of a single mesonic field (of mass dimension one) is not known.  We have assumed \cite{Fariborz:2015vsa,Fariborz:2019zht} that this relationship is of a simple form where the underlying composite fields of quarks inside a scalar meson are linearly proportional to the meson field.   If this assumption is a good approximation to the exact relationship between the meson fields and their underlying quark fields,  then the scale factor adjusting the mass dimensions should reflect certain  characteristics of the meson.   Chiral symmetry requires that all members of the same chiral nonet have the same scale factor  \cite{Fariborz:2015vsa,Fariborz:2019zht} (i.e., the universality condition), which is examined in this work, in testing the proposed bridge,  by independently computing the scale factors for the $K_0^*$  isodoublet scalar system and comparing with the previous computation of these factors for the $a_0$  isovector system of 
 Refs.~\cite{Fariborz:2015vsa,Fariborz:2019zht}.

At the mesonic level, our framework is the generalized linear sigma model of \cite{GLSM} that we use  to demonstrate the bridge between chiral Lagrangians and QCD sum-rules.   This  framework is formulated in terms of  a quark-antiquark chiral nonet and a four-quark chiral nonet, and even though there are no direct connections to the underlying quark world,   the distinction between two and four-quark nonets is  made through the  U(1)$_{\rm A}$ anomaly.     It is shown in 
\cite{GLSM} (and references therein) how the framework can incorporate various low-energy experimental data to disentangle two- from four-quark components of each members of the scalar meson nonet.   While this information is valuable it is not complete.     
Several four-quark composite fields, each with the same overall quantum numbers of a given scalar meson, can be formed out
of different combinations of color and spin (see for example \cite{GLSM_inst}),   but these combinations  cannot be disentangled solely on the basis of chiral symmetry -- a limitation of chiral Lagrangians (such as those of  \cite{GLSM}).
QCD sum-rules, on the other hand, have their own limitations ---   although they directly utilize the specific quark currents,  but when probing a complicated
scalar meson substructure for which there are numerous
possibilities for mixing among two- and four-quark currents, the disentanglement of two- from four-quarks 
is difficult to achieve in a self contained manner within its framework.  Addressing these limitations are  examples of the mutual benefits that this bridge provides: The disentanglement of two- from four-quark currents that can be determined at the mesonic level  can enhance (and simplify) the overall analysis of QCD sum-rules, 
and reciprocally,  the self consistency
checks within the QCD sum-rules can favor one combination of four-quark currents versus the other and remedy a
gap in the chiral Lagrangian approach which, due to the
lack of direct connection to the underlying quark fields,
is oblivious to various four-quark currents. 
 Establishing an interplay between chiral Lagrangians and QCD sum-rules has been the centerpiece of our proposal in Refs.~\cite{Fariborz:2015vsa,Fariborz:2019zht}.    This idea is not limited to the scalar channel and/or a specific type of chiral Lagrangian or a particular  variant of QCD sum-rules.

\section{Methodology: Scale Factor Matrices}
We begin by defining our notation. At the mesonic level,   we employ the generalized linear sigma model of \cite{GLSM} which is formulated in terms of two chiral nonets  $M$ and $M'$ that respectively represent a quark-antiquark nonet and a four-quark nonet (a ``molecule'' type and/or a diquark-antidiquark type) underlying substructure.      Both chiral nonets transform in the same way under chiral transformation but differently under U(1)$_{\rm A}$:
\begin{eqnarray}
M & \rightarrow& U_L \,  M \,  U_R^\dagger,  \, M\rightarrow e^{2i\nu}M\nonumber \\
M' & \rightarrow& U_L \,  M' \,  U_R^\dagger, \,  M'\rightarrow e^{-4i\nu}M'
\label{M_trans}
\end{eqnarray}
The axial charge is the main tool for distinguishing these two nonets.   Each of these two chiral nonets can be expressed in terms of a
scalar and pseudoscalar meson nonet
\begin{eqnarray}
M & = & S + i\phi \nonumber \\
M' & = & S' + i \phi'
\label{M_SP}
\end{eqnarray}
where the two scalar meson nonets contain the two- and four-quark ``bare'' (unmixed) scalars  
\begin{equation}
S=
\begin{pmatrix}
S_1^1 & a_0^+ & \kappa^+  \\
a_0^- & S_2^2 & \kappa^0 \\
\kappa^- & {\bar \kappa}^0 & S_3^3
\end{pmatrix}, 
\hskip 0.5cm
S' =
\begin{pmatrix}
{S'}_1^1 & {a'}_0^+ & {\kappa'}^+  \\
{a'}_0^- & {S'}_2^2 & {\kappa'}^0 \\
{\kappa'}^- & {\bar {\kappa'}}^0 & {S'}_3^3 \\
\end{pmatrix}
\label{SpMES}
\end{equation}
and similar matrices for $\phi$ and $\phi'$.   The framework of Ref.~\cite{GLSM,GLSM_inst} provides a detailed analysis of the mixing between these two ``bare'' nonets and how that results in a description of mass spectrum, decay widths and scattering analysis of scalar as well as pseudoscalar mesons below and above 1 GeV.    In this picture and of specific interest in this work,  the physical isodoublet scalars $K_0^*(700)$ and $K_0^*(1430)$ become a linear admixture of two- and  four-quark components $\kappa$ and $\kappa'$ respectively.
Understanding the physical characteristics of $K_0^*(700)$, particularly its substructure,  has posed many challenges and has resulted in numerous  investigations \cite{PDG}.  Particularly, the  possibility of a non-quark-antiquark nature of this state has been extensively studied 
\cite{Pelaez_kappa,GLSM,NLCL_kappa,Oller:1998zr,Jamin:2000wn,Oller:2003vf,Oller:2004xm,Guo:2012yt,Giacosa:2006tf}.

The transformation properties (\ref{M_trans}) as well as the decompositions (\ref{M_SP}) are direct consequences of the assumed underlying quark configurations.   The two mesonic-level  chiral nonets  $M$ and $M'$ can be mapped to the quark-level chiral nonets $\m$ and $M'_{\rm QCD}$.   For example, Eq.~(\ref{M_trans}) implies
\begin{equation}
(\m)_a^b \propto ({\bar q}_R)^b ({q_L})_a  \Rightarrow \left(S_{\rm QCD}\right)_a^b \propto q_a(x) {\bar q}^b(x)~,
\end{equation}
where $a$ and $b$ are flavor indices and each can take values of 1 to 3.
To make the exact connection to the quark world we need to make a specific choice for the proportionality factor, and with no loss of generality we choose $\left(S_{\rm QCD}\right)_a^b  = q_a(x) {\bar q}^b(x)$.  The local composite operator $S_{QCD}$ thus provides a current that is a necessary entity in QCD sum-rule methodology \cite{SVZ}.     Similarly,  $M'_{\rm QCD}$ can be  mapped to quark-level composite field configurations.   However, in this case there are several options,  each representing a different angular momentum, spin, flavor and color configurations for diquark-antidiquark combination.   Here we do not list such quark configurations and the specific form used for our analysis will be given below.

We assume a simple relationship between the quark-level nonets $M_{\rm QCD}$ and ${M'}_{\rm QCD}$ and the physical mesonic-level nonets $M$ and $M'$ via a scale-factor matrix that adjusts the mass dimensions
\begin{equation}
M=I_{M} M_{\rm QCD}\,,~
M'=I_{M'} {M'}_{\rm QCD}~.
\label{Mp_scale}
\end{equation}
As shown in \cite{Fariborz:2015vsa,Fariborz:2019zht}, chiral symmetry imposes the following constraints on the scale factor matrices
\begin{gather}
 [U_R,I_{M}]= [U_L,I_{M}]=0~,
  \label{I_M}\\
 [U_R,I_{M'}]= [U_L,I_{M'}]=0~, 
 \label{I_Mp}
\end{gather}
implying that the scale matrices are multiples of the identity matrix
\begin{equation}
I_M=-\frac{m_q}{\Lambda^3}\times \mathds{1}\,,~I_{M'}=\frac{1}{{\Lambda'}^5}\times  \mathds{1}\, ,
\label{scale_factors}
\end{equation}
where the (constant) scale factor quantities $\Lambda$ and $\Lambda'$ have dimensions of energy  that must be determined, and the quark mass factor $m_q=(m_u+m_d)/2$ has been chosen to  result in renormalization-group invariant currents as discussed below.  
This methodology can be generalized to include additional substructures (e.g., glueball components) through an additional scale factor.  
The scale factors $\Lambda$ and $\Lambda'$ have been determined in the study of $a_0$ isotriplet states \cite{Fariborz:2015vsa,Fariborz:2019zht} 
and will be redetermined here for isodoublet system to demonstrate universality.  

We now consider the specific example of
the isodoublets $K_0^*(700)$ and $K_0^*(1430)$, for which the physical states are related to the QCD operators via
\begin{equation}
{\bf K}=
\begin{pmatrix}
K_0^*(700)\\
K_0^*(1430)
\end{pmatrix}
= L_\kappa^{-1}
\begin{pmatrix}
S^3_2\\
\left(S'\right)^3_2
\end{pmatrix}
=
L_\kappa^{-1} I_\kappa J^{\rm QCD} 
\label{K_def}
\end{equation}
where $L^{-1}_\kappa$ is the rotation matrix that disentangles two- from four-quark components of isodoublets, $I_\kappa$    is formed out of the scale factors defined for the two chiral nonets in \eqref{scale_factors}, and  $J^{{\rm QCD}}$  is constructed from two- and four-quark operators (the specific form will be given below):
\begin{equation}
L^{-1}_\kappa=\begin{pmatrix}
\cos\theta_\kappa & -\sin\theta_\kappa
\\
\sin\theta_\kappa & \cos\theta_\kappa
\end{pmatrix}
\,,~I_\kappa =
\begin{pmatrix}
\frac{-m_q}{\Lambda^3} &0 \\
0 & \frac{1}{{{\Lambda'}^5}}
\end{pmatrix}
~.
\label{I_matrices}
\end{equation}
Since \eqref{K_def} relates the physical states to QCD operators, we define the 
projected physical currents
$J^P = L_\kappa^{-1} I_\kappa J^{\rm QCD}$ that define a
 physical correlation function matrix  $\Pi^P$  constructed from a physically-projected QCD correlation function matrix  $\Pi^{\rm QCD}$
\begin{gather}
 \Pi^P(Q^2) = {\widetilde {\cal T}}^\kappa \Pi^{\rm QCD}(Q^2)  {\cal T}^\kappa\,,~~{\cal T}^\kappa= I_\kappa \, L_\kappa
 \label{phys_corr}
 \\
 \Pi^{\rm QCD}_{mn}(x) =\langle 0| {\rm T}  \left[ J^{\rm QCD}_m (x) J_n^{\rm QCD}(0)^\dagger \right] |0 \rangle
\end{gather}
 where ${\widetilde {\cal T}^\kappa}$  denotes the transpose of the matrix ${\cal T}^\kappa$, and $m$ and $n$ can each take values of 1 (for quark-antiquark current) and 2 (for diquark-antidiquark current) as defined below.

The projected physical correlator matrix is diagonal, providing a self-consistency condition between elements of the QCD correlation function matrix.  In our  $2\times 2$ $K_0^*$ isodoublet  system we have the following constraint from the vanishing of off-diagonal elements. We note that a minor typographical error in Ref.~\cite{Fariborz:2015vsa} is corrected in \eqref{constraint} and \cite{Fariborz:2019zht}.
 \begin{equation}
 \Pi_{12}^{\rm QCD} = -
\left[
{
 \frac{    {\widetilde {\cal T}}^\kappa_{11} \Pi_{11}^{\rm QCD} {\cal T}^\kappa_{12}
    + {\widetilde {\cal T}}^\kappa_{12} \Pi_{22}^{\rm QCD} {\cal T}^\kappa_{22}
 }
 {{\widetilde {\cal T}}^\kappa_{11}  {\cal T}^\kappa_{22} + {\widetilde {\cal T}}^\kappa_{12}  {\cal T}^\kappa_{12}
 }
}
\right]~.
\label{constraint}
\end{equation}
The relation \eqref{constraint} 
will be used as input for $ \Pi_{12}^{\rm QCD}$
because the QCD off-diagonal correlator is unknown and not readily calculable because of its complicated higher-loop topology.

\section{QCD Sum-Rule Analysis of Scale Factors} 
QCD sum-rule methodologies are based on quark-hadron duality, and apply an integral transform  to a dispersion relation relating the QCD  and hadronic contributions  to the projected physical correlators  \cite{SVZ,Reinders:1984sr}.  
The mixing  matrix $L_\kappa$ must disentangle individual states so  a sum-rule method is needed to check whether a residual effect of multiple states is occurring because of an insufficiently accurate mixing matrix. 
Laplace sum-rules are not suitable because they suppress heavier states, so 
Gaussian sum-rules will be employed  because they provide similar weight to all states \cite{gauss,harnett_quark}.  The hadronic part of the Gaussian sum-rule is given by 
\begin{equation}
G^{H} ({\hat s}, \tau) =\frac{1}{\sqrt{4\pi\tau}}
\int\limits_{s_{th}}^{\infty} \!\! dt \,{\rm exp} \left[  {\frac{-({\hat s} - t)^2}{4\tau}}\right]\,\rho^H(t)~.
\label{GSR}
\end{equation}
 The hadronic spectral function $\rho^H$  in \eqref{GSR} is determined from the mesonic fields and a QCD continuum  above the  continuum threshold $s_0$:
\begin{eqnarray}
\rho^H(t)&=&
\frac{1}{\pi} {\rm Im} \Pi^{ H}(t) +\theta\left(t-s_0\right){\frac{1}{\pi}} {\rm Im} \Pi^{ \rm QCD}(t)
\label{spectral}
 \\
\Pi^{ H}_{ij} \left(q^2\right)&=&\int d^4x\, e^{iq\cdot x}
\langle 0| {\rm T} \left[ {\bf K}_i (x) {\bf K}^\dagger_j(0) \right] |0 \rangle
\nonumber \\
&=&\delta_{ij} \,
\left(
\frac{1}{{m_{\kappa i}^2 - q^2 -i m_{\kappa i}\Gamma_{\kappa i}}}
\right)~.
\label{PiH}
\end{eqnarray}
The last term in \eqref{spectral}  represents the QCD continuum contribution inherent in QCD sum-rule methods \cite{SVZ,Reinders:1984sr}. 
The effect of final state interactions in the
$\pi K$ channel is quite large near the  kappa pole.   Within the framework of generalized linear sigma model, these are estimated in \cite{GLSM_piK}.
 
The hadronic and QCD contributions to the Gaussian sum-rules are now equated:
\begin{equation}
G^H\left(\hat s, \tau\right)
=\widetilde{\cal T}^\kappa G^{\rm QCD}\left(\hat s,\tau ,s_0\right){\cal T}^\kappa
\label{full_GSR}
\end{equation}
where the QCD continuum has been absorbed from the hadronic side into the QCD contributions.  Methods for calculating the QCD prediction $G^{\rm QCD}$ from the underlying correlation function are reviewed in \cite{harnett_quark}.  
The QCD side  of \eqref{full_GSR} is diagonalized via the constraint \eqref{constraint}  and the hadronic side of \eqref{full_GSR} is diagonal  because the rotation matrix disentangles  the states  
\begin{equation}
G^H =
\begin{pmatrix}
{(G^H)}_{11} & 0 \\
0 & {(G^H)}_{22}
\end{pmatrix}
=\widetilde{\cal T}^\kappa G^{\rm QCD}{\cal T}^\kappa\,.
\label{G_matrix}
\end{equation}
The resulting diagonal elements of \eqref{G_matrix} are given by 
{\allowdisplaybreaks
\begin{gather}
G_{11}^H(\hat s,\tau)=a A
G_{11}^{\rm QCD}\left(\hat s,\tau, s_0^{(1)}\right)-bB
G_{22}^{\rm QCD}\left(\hat s,\tau, s_0^{(1)}\right)
\label{G_eqs}
\\
G_{22}^H(\hat s,\tau)=-aB
G_{11}^{\rm QCD}\left(\hat s,\tau, s_0^{(2)}\right)+bA
G_{22}^{\rm QCD}\left(\hat s,\tau, s_0^{(2)}\right)
\nonumber
\\
A=\frac{\cos^2\theta_\kappa}{\cos^2\theta_\kappa-\sin^2\theta_\kappa}\,,~
B=\frac{\sin^2\theta_\kappa}{\cos^2\theta_\kappa-\sin^2\theta_\kappa}
\\
a=\frac{m_q^2}{\Lambda^6}\,,~b=\frac{1}{\left(\Lambda'\right)^{10}}
\end{gather}
}where $G_{11}^H$ and $G_{22}^H$ respectively represent $K_0^*(700)$ and  $K^*_0(1430)$ contributions, and the factor of $m_q^2$ is combined with  $G_{11}^{QCD}$  for renormalization-group purposes.  Note that each sum-rule has its own continuum threshold represented by $s_0^{(1)}$ and $s_0^{(2)}$, and  the constraint \eqref{constraint} has been used within the QCD prediction. 

The scale factors $\Lambda$  and $\Lambda'$ for the isodoublet $K_0^*$ scalar meson system can now be calculated.  The QCD currents in \eqref{K_def} are
\begin{gather}
J^{\rm QCD}=\begin{pmatrix}
J_1\\
J_2
\end{pmatrix}
\,,~
J_1=\bar ds \\
\begin{split}
J_2&=\sin(\phi) u^T_\alpha C\gamma_\mu\gamma_5 s_\beta\left(\bar d_\alpha\gamma^\mu\gamma_5 C\bar u_\beta^T-\alpha\leftrightarrow \beta \right)
\\
&
+\cos(\phi) d^T_\alpha C\gamma_\mu s_\beta\left(\bar d_\alpha\gamma^\mu C\bar u_\beta^T+\alpha\leftrightarrow \beta \right)
\end{split}
\end{gather}
where $C$ is the charge conjugation operator and $\cot\phi=1/\sqrt{2}$ \cite{Chen:2007xr}.
Given an input of $\cos\theta_\kappa=0.4161$ from chiral Lagrangians \cite{GLSM} and the physical mass and width of the $K_0^*$ states (we use $m_\kappa=824 \,{\rm MeV}$, $\Gamma_\kappa=478\,{\rm MeV}$ for the $K_0^*(700)$ and $m_K=1425 \,{\rm MeV}$, $\Gamma_K=270\,{\rm MeV}$ for the $K_0^*(1430)$ to be consistent with \cite{PDG}) one can solve \eqref{G_eqs}  for the (constant) scale factors $\Lambda$ and $\Lambda'$, and optimize  the continuum thresholds to minimize the $\hat s$ dependence of the scale factors.  

The correlation function for  the two-quark current $J_1$ is
given in \cite{Zhang:2009qb,Du:2004ki} and the methods of \cite{harnett_quark} can then be used to form the Gaussian sum-rule:
{\allowdisplaybreaks
\begin{gather}
\begin{split}
&G_{11}^{\rm QCD}\left(\hat s, \tau,s_0\right)=\\
&\frac{3}{8\pi^2}
\int\limits_0^{s_0} \!\!
t\,dt\left[\left(1+\frac{17}{3}\frac{\alpha_s}{\pi}\right)
-2\frac{\alpha_s}{\pi}\log{\left(\frac{t}{\sqrt{\tau}}\right)}
\right]W\left(t,\hat s,\tau\right)
\\
&   +\frac{\pi n_c\rho_c^2 }{m_s^*m_q^*}
\int\limits_0^{s_0}
t J_1\left(\rho_c\sqrt{t}\right) Y_1\left(\rho_c\sqrt{t}\right) 
W\left(t,\hat s,\tau\right)\,dt
\\
& + \exp{\left(-\frac{\hat s^2}{4\tau}\right)}\left[
\frac{1}{2\sqrt{\pi\tau}}\left\langle C^s_4{\cal O}^s_4\right\rangle-
\frac{\hat s}{4\tau\sqrt{\pi\tau}}\left\langle C^s_6{\cal O}^s_6\right\rangle
\right]~,
\end{split}
\label{gauss_scalar_QCD}
\\
W\left(t,\hat s,\tau\right)=\frac{1}{\sqrt{4\pi\tau}}\exp{\left(-\frac{\left(t-\hat s\right)^2}{4\tau}\right)}
\\
\left\langle C_4^s {\cal O}_4^s\right\rangle=
 \left\langle m_s \overline{q}q\right\rangle 
+\frac{1}{2} \left\langle m_s \overline{s}s\right\rangle +
\frac{1}{8\pi} \left\langle\alpha_s G^2 \right\rangle\,,~
\label{c4_scalar}
\\
\begin{split}
\langle C^s_6{\cal O}^s_6\rangle&=
-\frac{1}{2}\left\langle m_s\overline{q}\sigma G q\right\rangle
-\frac{1}{2}\left\langle m_q\overline{s}\sigma G s\right\rangle
\\
&\!\!\!-\frac{16\pi}{27}\alpha_s
\left( \left\langle \bar q q\right\rangle^2+\left\langle \bar s s\right\rangle^2  \right)
-\frac{48}{9}\alpha_s\left\langle \bar q q\right\rangle\left\langle \bar s s\right\rangle
\,,
\end{split}
\label{c6_scalar}
\end{gather}
}where $q$ denotes the non-strange $u,d$ quarks and vacuum saturation has been used  for the dimension-six (four-quark) condensates.
Because 
$G_{11}^{\rm QCD}$ is being combined with  $m_q^2$, the combination satisfies a homogenous renormalization-group equation, which requires evaluating all running quantities
at the renormalization scale $\nu^2=\sqrt{\tau}$ \cite{harnett_quark}. 
Similarly, the Gaussian sum-rule related  to the four-quark current $J_2$ is obtained via 
\begin{gather}
G_{22}^{\rm QCD}\left(\hat s, \tau,s_0\right)=
\int\limits_{0}^{s_0}\!\!dt\,   W\left(t,\hat s,\tau\right)
\, \rho^{\rm QCD}(t) 
\end{gather}
where $\rho^{\rm QCD}(t)$ is given in \cite{Chen:2007xr}.
Because this result is leading-order, 
renormalization of the current $J_2$ represents a higher-order effect and the Gaussian sum-rule $G_{22}^{\rm QCD}$ effectively satisfies a homogenous renormalization-group equation, allowing application of the renormalization-group results of Ref.~\cite{harnett_quark}.
For the QCD input parameters we use PDG values \cite{PDG} (quark masses, and $\alpha_s$)
 and the following QCD condensate \cite{Reinders:1984sr,Narison:2011rn,Beneke:1992ba,Belyaev:1982sa} and instanton liquid model parameters \cite{Shuryak:1982qx,Schafer:1996wv}
{\allowdisplaybreaks
\begin{gather}
 \langle\alpha_s G^2\rangle
      = (0.07\pm 0.02)\, {\rm GeV^4} \,,
      \label{GG}
      \\
\frac{\left\langle \overline{q}\sigma G q\right\rangle}{\langle \bar q q\rangle}=\frac{\left\langle \overline{s}\sigma G s\right\rangle}{\langle \bar s s\rangle}=(0.8\pm 0.1) \,{\rm GeV^2}
\label{mix}
\\
\langle \bar q q\rangle=-\left(0.24\pm 0.2 \,{\rm GeV}\right)^3\,,~\langle \bar s s\rangle=(0.8\pm 0.1)\langle \bar q q\rangle
\label{O6}
\\
  n_{{c}} = 8.0\times 10^{-4}\ {\rm GeV^4}~,~\rho =1/600\,{\rm MeV}~,
  \label{inst}
  \\
  m^*_q=170\,{\rm MeV}\,,~m^*_s=220\,{\rm MeV}~.
\end{gather}
} 
The instanton parameters $\rho$ and $n_c$ have an estimated uncertainty of 15\% and the quark zero-mode effective masses $m^*$ are correlated with the uncertainty in $\rho$ and the quark condensate \cite{Shuryak:1981ff}. 
The $m_s/m_q=27.3$ ratio \cite{PDG} is of particular importance because it appears in both the QCD inputs and as a parameter in the chiral Lagrangian analysis. The effect of the theoretical uncertainty in the gluon condensate (the dominant QCD condensate effect) on the $a_0$ scale factors was examined in Ref.~\cite{Fariborz:2019zht} and found to be a small numerical effect.
We choose $\tau=3\,{\rm GeV^4}$ consistent with the central value used in Refs.~\cite{harnett_quark,Harnett:2000fy}.

\begin{figure}[htb]
\centering
\includegraphics[width=\columnwidth]{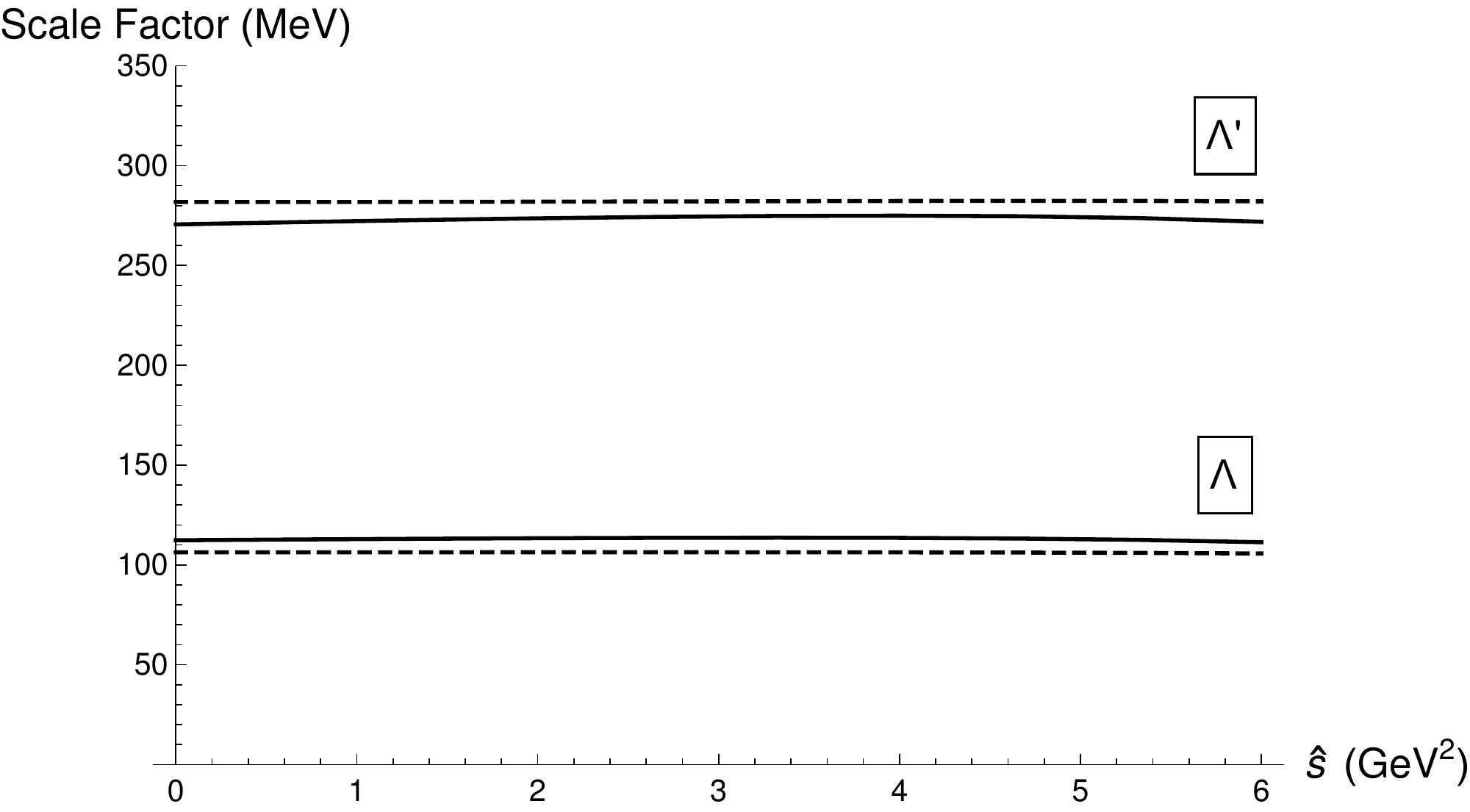}\hspace{0.02\columnwidth}
\caption{The scale factors $\Lambda$ (lower pair of curves) and $\Lambda'$ (upper pair of curves) are shown as a function of $\hat s$
for optimized continuum thresholds in Table \ref{scale_tab}.  Solid curves are for the $K_0^*$ channel and dashed curves are for the $a_0$ channel.
}
\label{scale_fig1}
\end{figure}

Fig.~\ref{scale_fig1} shows the $\hat s$ dependence of the scale factors for the optimized values of the continuum for both the $a_0$ \cite{Fariborz:2015vsa,Fariborz:2019zht} and $K_0^*$ channels.
The $a_0$ channel results have  been updated from \cite{Fariborz:2015vsa,Fariborz:2019zht} to use $\cos\theta_a=0.6304$ consistent with the $m_s/m_q$ mass ratio used for the $K_0^*$  analysis. 
The remarkable independence of the scale factors on the auxiliary sum-rule parameter $\hat s$ demonstrates the validity of the scale-factor matrices connecting chiral Lagrangians mesonic fields and the quark-level operators in QCD sum-rules. 
The small  energy ($\hat s$) dependence of the scale factors demonstrates that any additional dynamics needed to connect the chiral Lagrangian and QCD sum-rule frameworks is small.
As is evident from Fig.~\ref{scale_fig1} and Table~\ref{scale_tab}, the best-fit predictions of the scale factors   clearly demonstrate the crucial universality property required by chiral symmetry.   

It is interesting that the scale factors are on the order of magnitude of $\Lambda_{QCD}$, particularly considering the inclusion of the quark mass factor in \eqref{scale_factors}.  Although we cannot make a direct connection between the scale factors and $\Lambda_{QCD}$, \eqref{Mp_scale} connects mesonic fields with QCD operators and has an implicit connection to the approximate $\Lambda_{QCD}$  scale of hadronization.    

\section{Discussion}
In addition to providing a bridge between chiral Lagrangians and QCD sum-rules, the scale factors can also be related to the chiral Lagrangian vacuum expectations values via 
$\langle S_1^1\rangle=-m_q\langle\bar u u\rangle/\Lambda^3$ 
and
  $\langle {S'}_1^1\rangle\approx 1.31\langle \bar d d\rangle \langle \bar s s\rangle/\Lambda'^{\,5}$ (note that vacuum saturation effects are embedded in the 1.31 numerical factor).    
 The relation $\langle {S'}_1^1\rangle$ is approximate because it  depends on the renormalization scale and relies upon the vacuum saturation approximation.  
As discussed in \cite{Fariborz:2015vsa,Fariborz:2019zht}, the resulting agreement is excellent for   $\langle S_1^1\rangle$  and provides the approximate scale  for $\langle {S'}_1^1\rangle$. 

\begin{table}[htb]
\begin{tabular}{c|cccc}
\hline
\rule{0pt}{3ex}   
Channel &  $s_0^{(1)}$ & $s_0^{(2)}$ & $\Lambda$ & $\Lambda'$ 
\\[2pt]
\hline
$K_0^*$ &1.61 & 3.04 & 0.114 & 0.276 
\\
$a_0$ & 1.68 & 2.88 & 0.106 & 0.282 
\\
\hline
\end{tabular}
\caption{Values for the optimized scale factors $\Lambda,~\Lambda'$ and continuum thresholds 
$s_0^{(1)},~s_0^{(2)}$  for the $a_0$ and $K_0^*$ channels.
All quantities are in appropriate powers of GeV. 
}
\label{scale_tab}
\end{table}

In summary, it has been shown that chiral symmetry transformation properties require that the  scale factor matrices serving as a bridge connecting QCD sum-rules and chiral Lagrangians must contain universal scale factors for all sectors of the chiral nonets.  
The scale factors determined in this work for the $K_0^*$ system are in remarkable  agreement with the corresponding values previously found in the $a_0$ channel \cite{Fariborz:2015vsa,Fariborz:2019zht} (see Table~\ref{scale_tab}), providing a key demonstration of the  universality property. 
With evidence for universal scale factors now established, 
more complicated sectors of the scalar nonets can therefore be simplified by taking input of the universal scale factors from other channels in the nonet.  
This powerful synergy between chiral Lagrangians and QCD will enable future progress on  more challenging and controversial aspects of low-energy hadronic physics.  It is also interesting to study  application of Eq.~\eqref{Mp_scale} in higher spin systems such as in vector and axial-vector meson sectors \cite{Carter:1995zi,Ko:1994en,Parganlija:2012fy} and examine whether quark currents for vectors and axials scale with the same scale factors of Eq.~\eqref{Mp_scale} and give rise to the vectors and axials mesonic fields.

\begin{acknowledgments}
This research was generously supported in part by the SUNY Polytechnic Institute Research Seed Grant Program.
TGS is grateful for the hospitality of AHF and SUNY Polytechnic Institute while this work was initiated.  
TGS and JH are grateful for research funding from the Natural Sciences and Engineering Research Council of Canada (NSERC), and 
AHF is grateful for a 2019 Seed Grant,  and the support of College of Arts and Sciences, SUNY Polytechnic Institute.
We thank Research Computing at the University of Saskatchewan for computational resources. 
\end{acknowledgments}

\end{document}